\documentclass[times,twocolumn,final,authoryear]{elsarticle}

\usepackage{jasr}
\usepackage{framed,multirow}
\usepackage{amssymb}
\usepackage{latexsym}
\usepackage{xcolor}
\usepackage[citebordercolor=white]{hyperref}
\usepackage[hyphenbreaks]{breakurl}
\usepackage[switch]{lineno}

\journal{Advances in Space Research}

\begin{document}

\verso{Vesna Borka Jovanovi\'{c} \textit{et al}}

\begin{frontmatter}

\title{Spectral index distribution over radio lobes of 4C 14.11 using astrophysical data in FITS format}

\author[1]{Vesna \snm{Borka Jovanovi\'{c}}\corref{cor1}}
\cortext[cor1]{Corresponding author.}
\ead{vborka@vinca.rs}
\author[1]{Du\v{s}ko \snm{Borka}}
\ead{dusborka@vinca.rs}
\author[1]{Arsenije \snm{Arseni\'{c}}}
\ead{byebyenoob@gmail.com}
\author[2]{Predrag \snm{Jovanovi\'{c}}}
\ead{pjovanovic@aob.rs}

\address[1]{Department of Theoretical Physics and Condensed Matter Physics (020), Vin\v{c}a Institute of Nuclear Sciences - National Institute of the Republic of Serbia, University of Belgrade, P.O. Box 522, 11001 Belgrade, Serbia}
\address[2]{Astronomical Observatory, Volgina 7, P.O. Box 74, 11060 Belgrade, Serbia}

\received{}
\finalform{}
\accepted{}
\availableonline{}
\communicated{}

\begin{abstract}
The goal of this paper is to investigate the flux and spectral index distribution of FR II radio galaxy 4C 14.11. We focused on the distribution of flux and spectral indices over the lobes, as well as in their hot spots. For that purpose, we used publicly available observations of this radio galaxy given at 1450 and 8440 MHz. Particularly, we used Leahy's Atlas of radio-emitting double radio sources, Jodrell Bank Centre for Astrophysics in Manchester, as well as NASA/IPAC Extragalactic Database. We found that the non-thermal (synchrotron) radiation dominates over the areas of the lobes. Distinction between hot spots and rest of the lobes are much smaller in the spectral index than in the flux. We also found that over the inner parts of both lobes, spectral index $\alpha$ is flat in average and significantly higher than 1.2, indicating that 4C 14.11 is old AGN.
\end{abstract}


\end{frontmatter}


\section{Introduction}

Active Galactic Nuclei (AGNs) are bright compact regions at the centers of active galaxies which are driven by the accretion of infalling material onto the central supermassive black hole (SMBH) \citep[see e.g.][for a review]{jova12}. They were originally discovered by Carl Seyfert in the 1943 \citep{seyf43}. Only small minority of Galactic Nuclei (around 10-15~\%) are active \citep{bemp17}. Besides, radio galaxies show some significant difference in their spectra and have different optical-to-radio flux ratios, according to which they can be classified to radio-loud and radio-quiet AGNs (see e.g. \citet{jova11} and references therein). 

Double Radio Sources Associated with Galactic Nuclei (DRAGNs) are large-scale double radio sources produced by outflows (jets) that are launched by processes in AGN. DRAGNs produce radio emission throughout regions that are much larger than the host galaxies themselves \citep{leah93}. DRAGNs are formed when an AGN produces two oppositely-directed plasma outflows that contain cosmic ray electrons and magnetic fields. DRAGNs main characteristics are synchrotron radiation, jets and lobes \citep{leah93}. The environs of a galaxy are not a perfect vacuum. The jets do not flow freely away from the AGN, but must push their way through these external media. In more powerful sources, the jets usually remain relativistic (and supersonic) out to great distances from their host galaxies, to form the ''clasical'' double-lobed structures. The ends of the jets move outwards more slowly than material flows along the jet. In the less powerful sources, the jets are slowed down enough on the scale of the galaxy to become subsonic and turbulent. DRAGNs are characterized by bright hotspots near the end of each lobe. They are part of the lobes and they mark the point where the jet collides with the surface of the lobe, i.e. the ends of the jets. In case the jet is still supersonic at this point, it will cause a system of strong shockwaves. The resulting high-pressure region will be seen as the hotspot \citep{leah93}.

The Fanaroff and Riley gave a classification scheme based on the degree of correlation between the high and low surface brightness in the lobes of extragalactic radio sources with their radio luminosity \citep{fana74}. The classification is divided in FR I and FR II types, and it is used to separate low power and high power radio galaxies, respectively \citep{blan19}. Let us mention here that while the luminosity is measured in W, the radio luminosity is measured in W Hz$^{-1}$, thus avoiding to specify a bandwidth over which it is measured. Fanaroff-Riley Type I (FR I) radio galaxies (radio powers, $P_{rad} < 10^{24}$ W Hz$^{-1}$ at 1.4 GHz) are usually characterized with their symmetric prominent jets and lobes, they are associated with the low excitation emission line galaxies. High brightness regions are often observed with double-sided radio jets. They show extended structures which interact with the external environment at large distances (range of kpc scales) and often form spectacular tailed radio galaxies \citep{lal21}. Fanaroff-Riley Type II (FR II) radio galaxies (radio powers, $P_{rad} > 10^{24}$ W Hz$^{-1}$ at 1.4 GHz) are usually characterized with large diffuse, double lobes and bright hot spots. They are edge-brightened with often asymmetric radio jets. The lobes are the result of backflow from the hot spots \citep{lal21}. The FR I preferentially reside in dense environments, such as clusters and groups of galaxies, whereas FR II are found in less dense environments. The jets of both, FR I and FR II radio galaxies are relativistic, but jets in FR I decelerate closer to the core than jets in FR II due to the differences in their environments \citep{lal21}.

\begin{figure*}[ht!]
\centering
\includegraphics[width=0.42\textwidth]{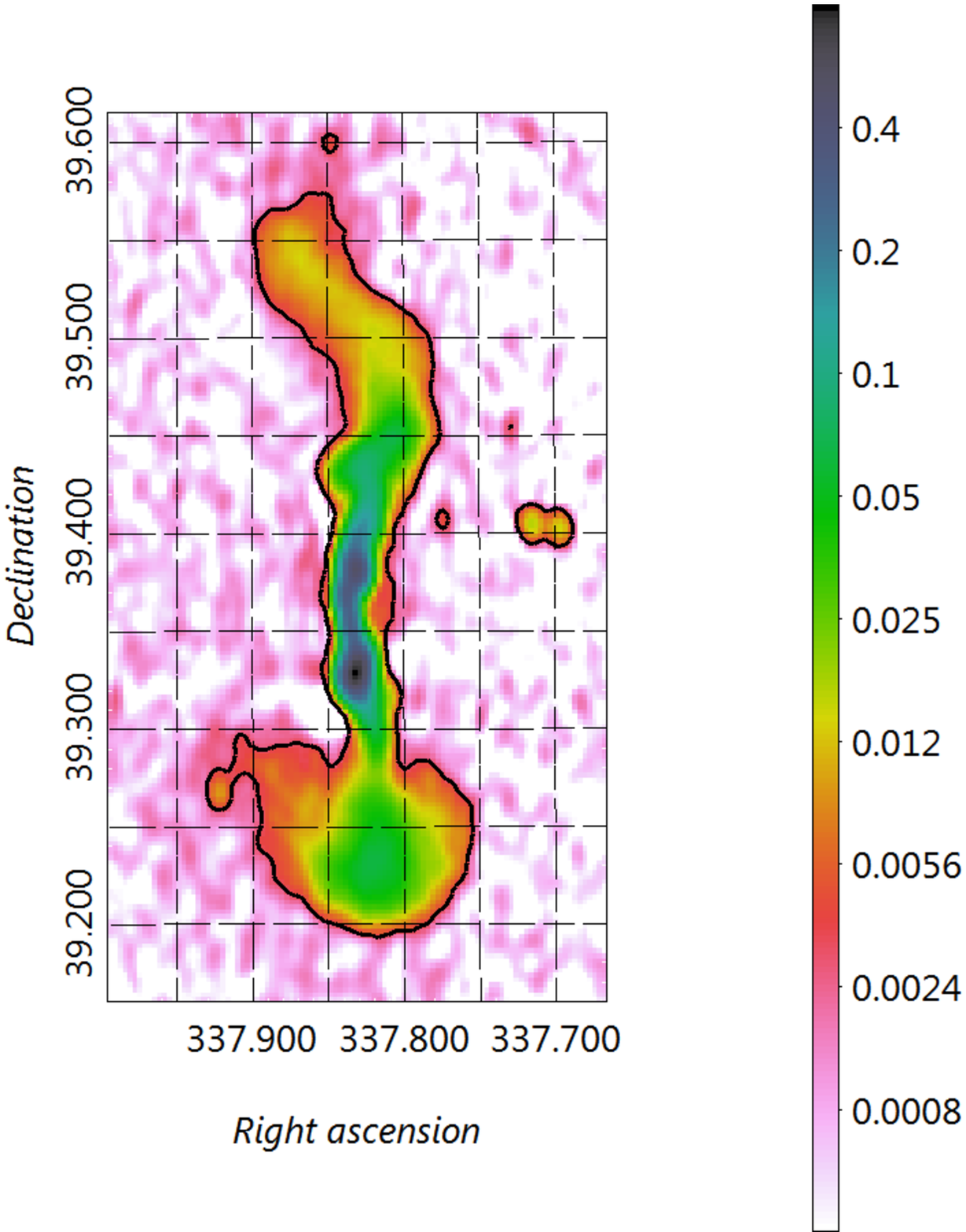}
\hfill
\includegraphics[width=0.56\textwidth]{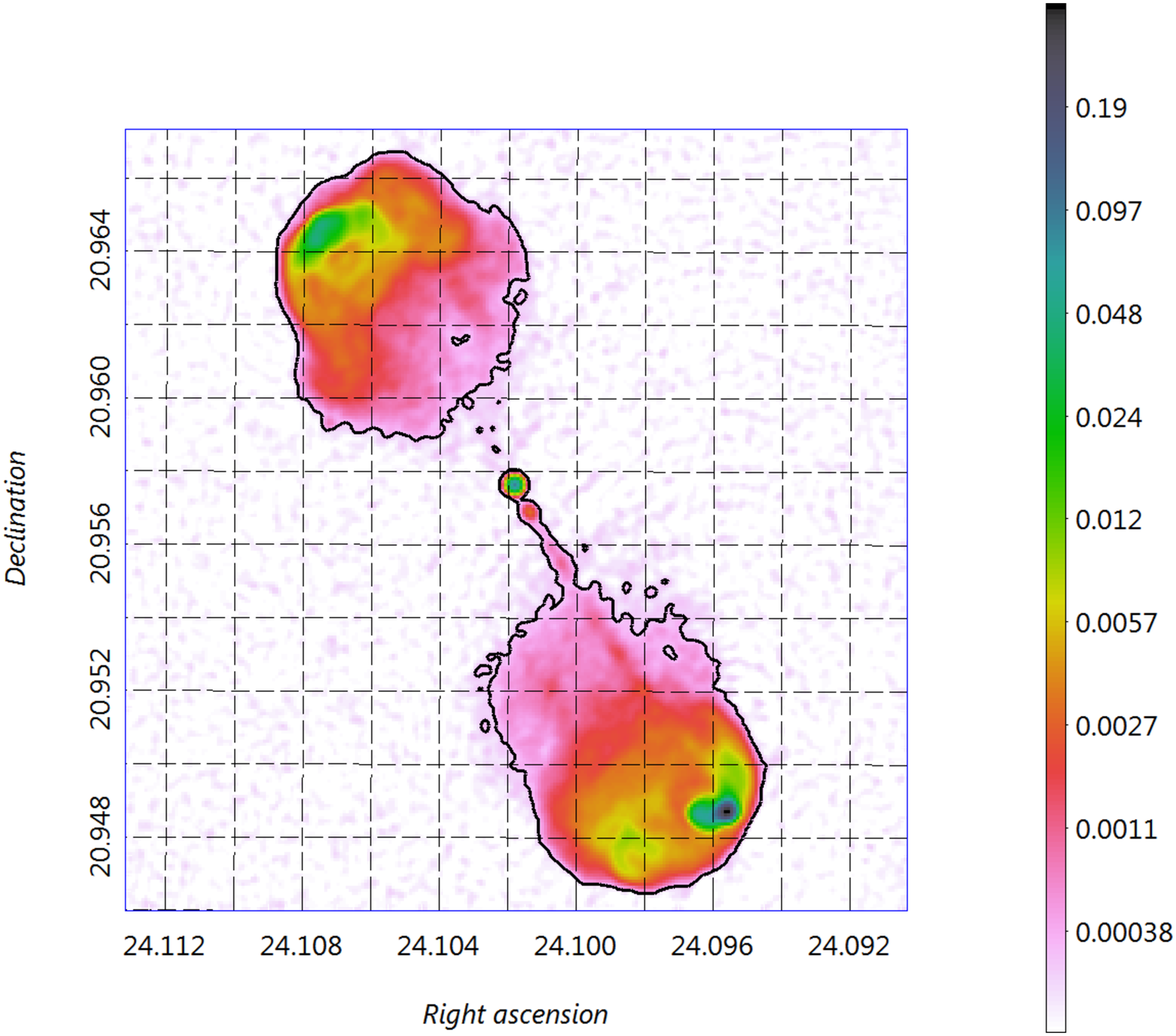}
\caption{3C 449 at $608$ MHz (left) and 3C 47 at $1650$ MHz (right), with flux density scales in Jy (1 Jy = $10^{-26}$ W m$^{-2}$ Hz$^{-1}$). Classic examples of FR I (left) and FR II (right) radio galaxies, mentioned in \citep{arse21}. FITS files taken from Leahy's Atlas.}
\label{fig01}
\end{figure*}

Astrophysical jets are fast, collimated outflows of plasma from compact celestial objects and serve as conduits for mass, momentum, energy and magnetic flux transport \citep{bemp17}. In case of the standard dynamical model of FR II radio galaxies, the relativistic plasma after being accelerated at the hotspots flows backwards towards the core. Positrons and electrons in the lobes of radio galaxies radiate by the synchrotron process and by inverse-Compton scattering against the cosmic microwave background (CMB) photons. In that way those particles lose energy \citep{kona13}.

Synchrotron radiation is the electromagnetic radiation emitted when relativistic charged particles are accelerated perpendicular to their velocity. It is produced artificially in some types of particle accelerators, or naturally by fast electrons moving through magnetic fields. The radiation produced in this way is polarized, where the direction of the electric vector depends on the direction of the local magnetic field. Also, it is produced over a wide range of energies: from the infrared region ($<$ 1 eV) to the hard X-ray region (100 keV or more). Most known cosmic radio sources emit synchrotron radiation. More about this radiation coming from the astrophysical objects: pulsars, supernova remnants, radio galaxies, and galaxy clusters, see in \citet{bott13}, and about radiation from supermassive black holes see in \citet{jova09}. It is often used to estimate the strength of large cosmic magnetic fields, as well as to analyze the contents of the interstellar and intergalactic media. In the case of radio lobes, synchrotron radiation has curved spectrum, and radio observations of such objects usually show the steepening of the spectrum at low frequencies. The steepening of the spectrum then interprets the age of electrons population that has generate the emission.

In a system of particles emitting synchrotron radiation under the effect of a magnetic field, the energy loss is higher for high-energy particles. The result is that, over time, the spectrum steepens, starting from the higher frequency end. If the steepening is particularly large, with spectral indices steeper than $\alpha$ = 1.2 \citep{komi94}, it is the signature of inactive or dying radio AGN where the central engine has stopped or its activity is substantially dimmed. If the frequency at which this ultra steepening occurs is lower, the time-span is larger (for a given magnetic field) and the remnant source is older \citep{morg21}. Radio spectral properties of a number of radio galaxies were discussed in e.g. \citet{hard96,hard97,brie20,leah91,perl93,khar08,fana21} and references therein.

In this paper we study the radio spectral index distribution of 4C 14.11 DRAGN. The paper is organized as follows: in \S2 we describe the used observations and our method for calculation of radio spectral index. In \S3 we present our results and discuss the spectral index distribution of 4C 14.11 radio source. Finally, in \S4 we point out the main conclusion in our study.

\begin{figure*}[ht!]
\centering
\includegraphics[width=0.48\textwidth]{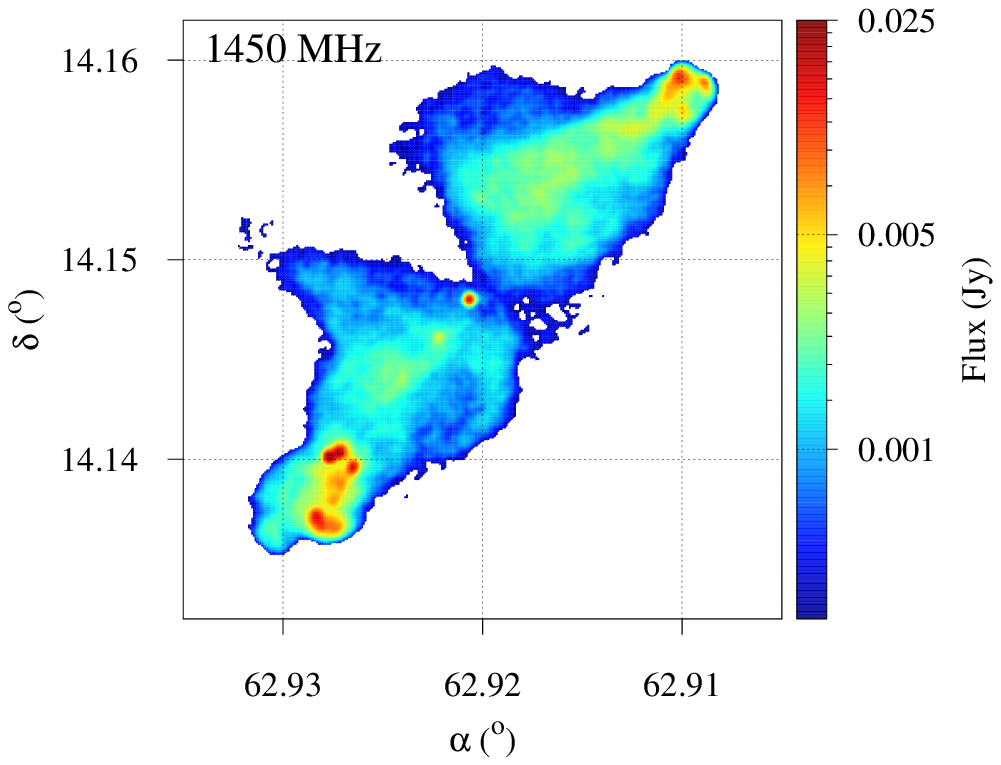}
\hfill
\includegraphics[width=0.48\textwidth]{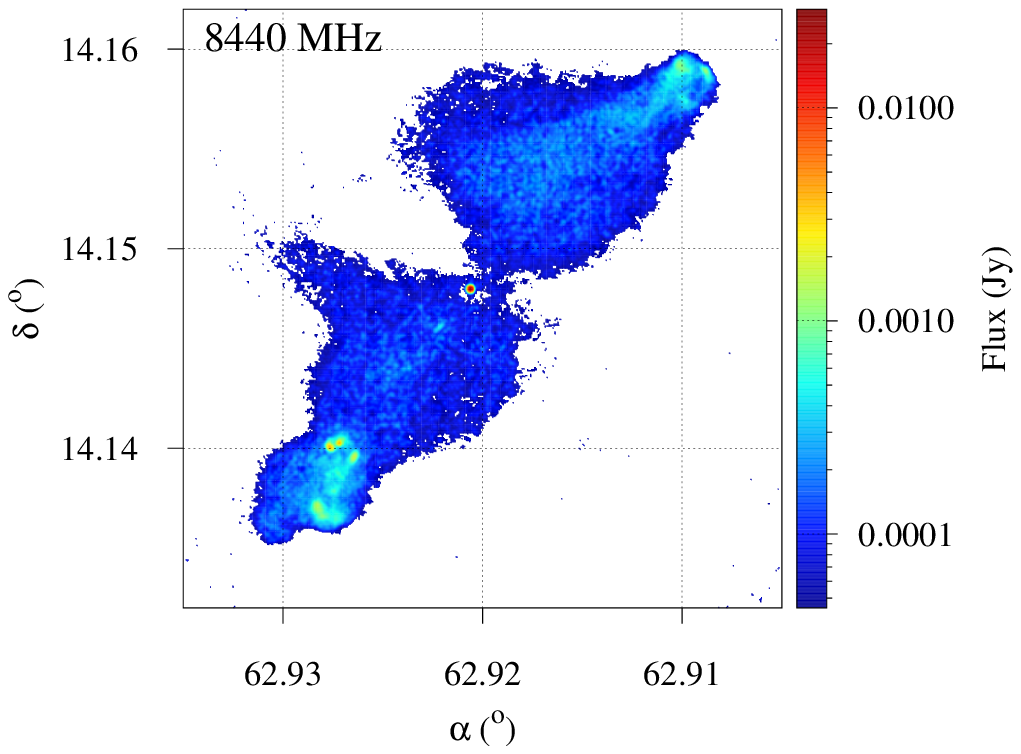}
\caption{The area of 4C 14.11 at 1450 MHz (left) and 8440 MHz (right), showing distribution of flux densities, while the surrounding is denoted with white color. The corresponding flux density scales are given (in Jy).}
\label{fig02}
\end{figure*}

\begin{figure*}[ht!]
\centering
\includegraphics[width=0.48\textwidth]{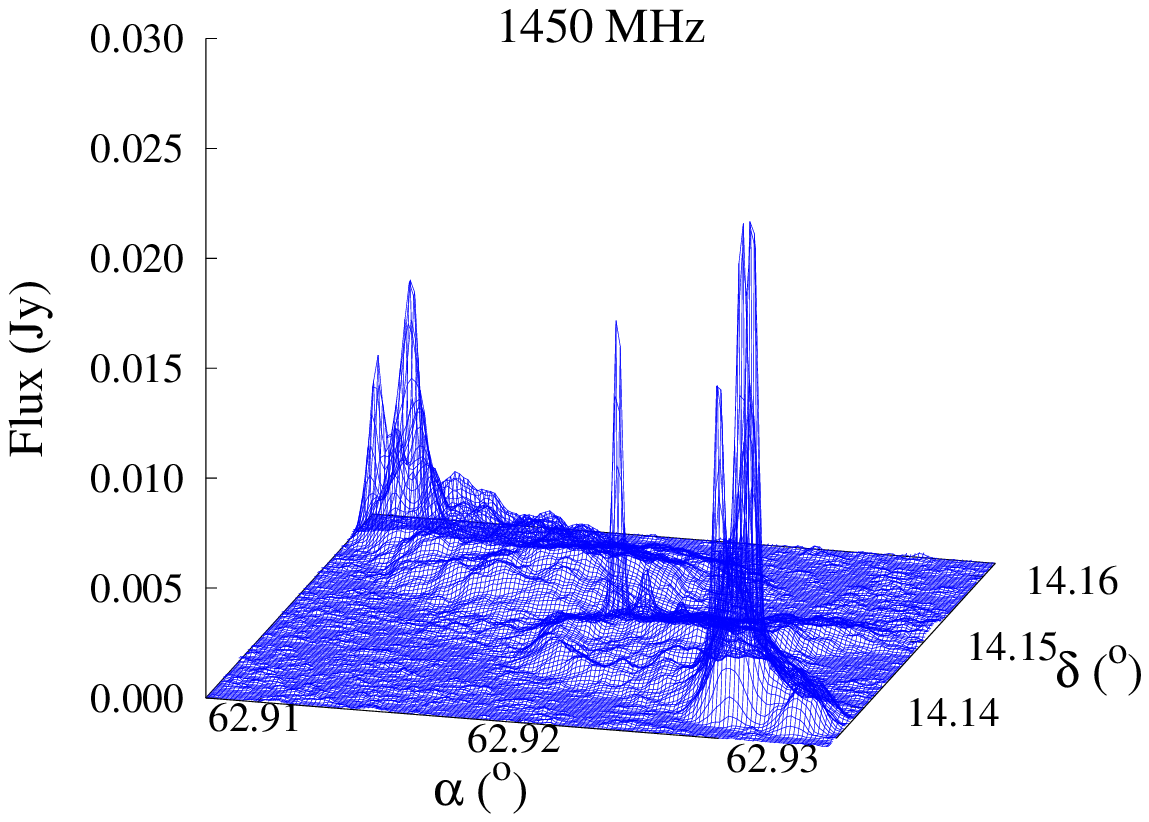}
\hfill
\includegraphics[width=0.48\textwidth]{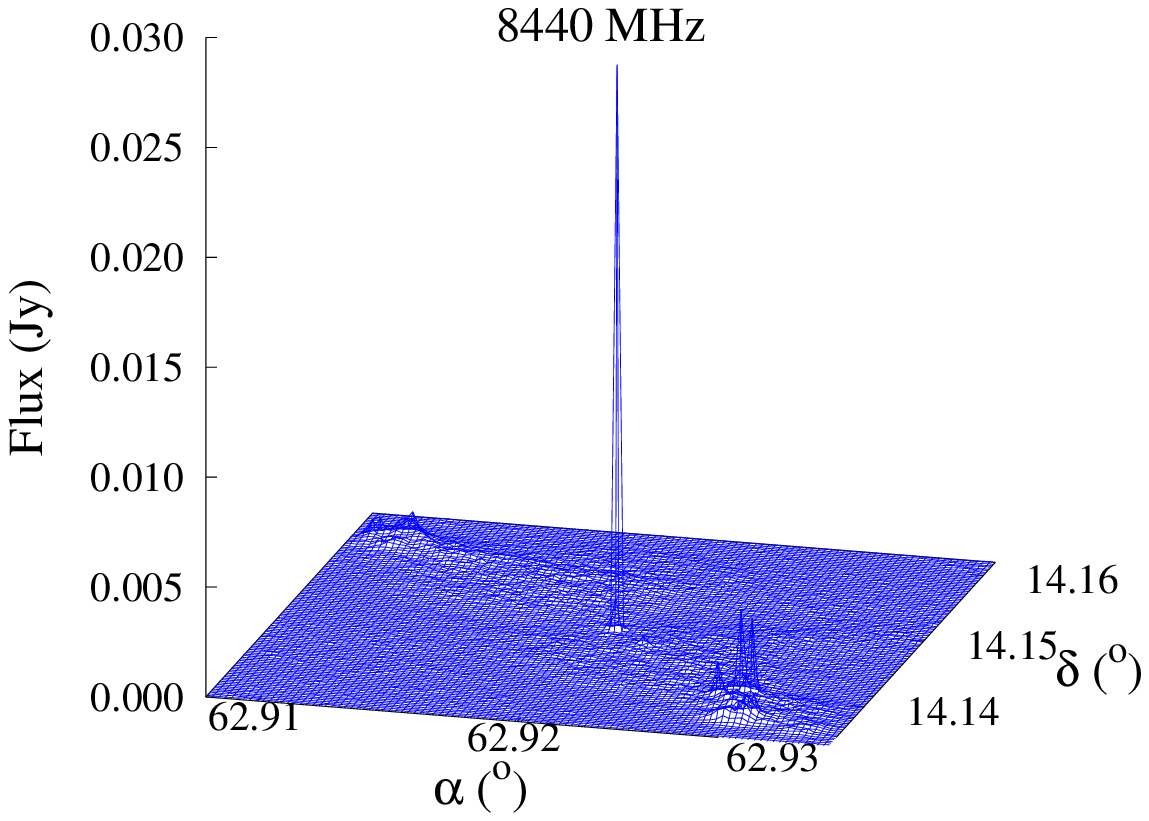}
\caption{The flux density 3D plot for 4C 14.11 at 1450 MHz (left) and 8440 MHz (right).}
\label{fig03}
\end{figure*}

\section{Data and method}

\subsection{FITS format of data}

FITS (short for Flexible Image Transport System) is the standard data format in astronomy that got popularized due to its flexibility and storage efficiency. First developed in the late 1970s, it was most commonly used for transferring images of astronomical observations, however it is not limited to this and can be used to store and transfer many other forms of astronomical data. Since 1982, it was endorsed by both NASA and the International Astronomical Union.

Structurally it consists of one or more sets of a Header and Data Units (called HDUs). The Header is human-readable and contains keywords and values which describe the data (such as position or time of observation), whereas the Data Units are packaged and form regularly spaced $n$-dimensional arrays (or $n$-cubes). Standard examples for these $n$-cubes would be a one dimensional spectrum or a two dimensional image where each coordinate corresponds to a value for flux density.

The FITS format has been designed with long term archiving in mind, so all versions of the FITS format are backwards-compatible, with the latest version being 4.0 (released in 2016.). For details see: \url{https://www.loc.gov/preservation/digital/formats/fdd/fdd000317.shtml}.

\subsection{Leahy's Atlas and NED database}

The sources for this study have been selected from Leahy's Atlas of DRAGNs (edited by \citet{leah13}) due to the availability of images of the listed radio galaxies at multiple frequencies as well as the wide range of research already conducted on plenty of these galaxies. The atlas itself was compiled by J.P. Leahy, A.H. Bridle and R.G. Strom, and consists of the $85$ closest DRAGNs contained in the 3CRR sample \citep{lain83}, with highly detailed images given at a variety of frequencies as well as basic information regarding each one, such as unique characteristics (\url{http://www.jb.man.ac.uk/atlas/}).

The actual FITS files for each source used in this study was found through the NASA/IPAC Extragalactic Database (NED). This is a continuously updated astronomical database hosted jointly by NASA and IPAC which collates information and bibliographic references regarding astronomical objects, thereby providing easily searchable and accessible data. More specifically, it also provides FITS files for publicly available images of astronomical objects from various surveys conducted over the years (such as Faint Images of the Radio Sky at Twenty-Centimeters (FIRST) or the Very Large Array Sky Survey (VLASS)). Database site: \url{http://ned.ipac.caltech.edu/}.

Some of the best known examples of FR I and FR II radio galaxies, mentioned in \citet{arse21}, with FITS data containing flux densities over the area of the source, are presented in Figure \ref{fig01}: 3C 449 at 608 MHz and 3C 47 at 1650 MHz.

\begin{figure*}[ht!]
\centering
\includegraphics[width=0.65\textwidth]{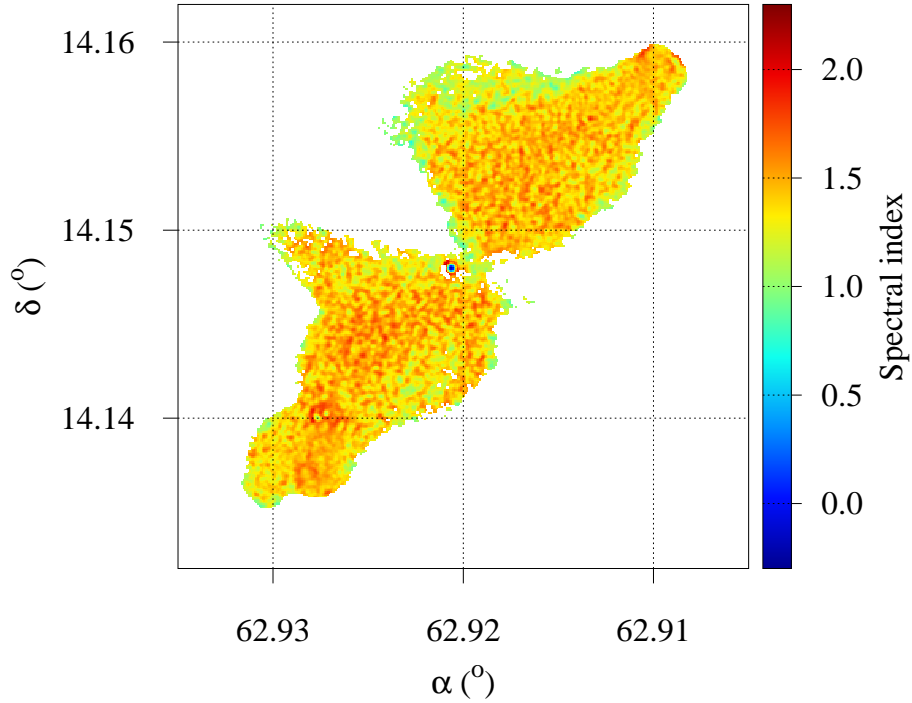}
\caption{The spectral index image of 4C 14.11 obtained combining 1450 MHz (21 cm) and 8440 MHz (3.6 cm) data.}
\label{fig04}
\end{figure*}

\subsection{Double radio source 4C 14.11}

The 4CR catalogue - 4$^{rd}$ Cambridge Catalog of Radio Sources, published in the 1960s, includes northern hemisphere radio sources detected at 178 MHz. There have been a series of Cambridge Catalogs of Radio Sources (the first from 1950 and the 10$^{th}$ in 2013). The early ones, including 3C and 4C were pioneering, and over the course of the series, various frequencies have been surveyed as new equipment is developed. References to entries in this catalogue use the prefix 4C followed by the declination in degrees, followed by a period, and then followed by the source number on that declination strip. 4C 14.11 (cross-identifications: 4C +14.11; PKS 0411+14; WISEA J041429.81+141623.6; 2MASS J04142982+1416231; LEDA 2819197) was added to the 3CRR sample by \citet{lain83} as its flux in the more accurate 4CT survey was high enough that it should have been included in 3CR. 

The method of calculation, that we have developed, first time is published in \citet{bork07} (where we investigated main Galactic radio-continuum loops I-VI), and explained in detail in \citet{bork12a}. Regarding our study of the properties of some radio source, as well as the calculation of its radio spectral index, see also the following papers: \citet{bork08} -- for Galactic loops V and VI, \citet{bork09} -- Monoceros loop, \citet{bork11} -- Cygnus loop, \citet{bork12b} -- 3C 349, \citet{bork12c} -- HB21,\citet{bork17} -- Lupus loop.

We showed that the method for defining a radio loop border and for determining the values of the brightness temperature and surface brightness, which we developed for the main Galactic Loops, could be applicable (and also rather efficient) to angularly large supernova remnants (SNRs). Then, we extended the same method to extragalactic radio sources. While in case of Galactic Loops and SNRs we had brightness temperature data, for objects in Leahy's Atlas we have available flux densities, but because of their linear dependence the same procedure can be used. Beside this, an extension of this method to much smaller objects is possible because the objects from Leahy's Atlas are observed with much better resolution.

Note that the only difference is that when calculating the flux density of DRAGNs, we do not need to consider background radiation as this has already been subtracted from observed data. With aim to distinguish the source from the surrounding objects, it is very important to determine the exact borders, i.e. minimum and maximum flux densities. For 4C 14.11 we found the following: $(S_{\nu,\mathrm{min}},S_{\nu,\mathrm{max}})$ = (0.00028 Jy, 0.025 Jy) at 1450 MHz (21 cm) and $(S_{\nu,\mathrm{min}},S_{\nu,\mathrm{max}})$ = (0.000045 Jy, 0.029 Jy) at 8440 MHz (3.6 cm).

Here we want just to mention, briefly, that the area of the investigated DRAGN was determined in three ways \citep{bork12a} - which is really a good way to check if our results are good:
\begin{enumerate}[(1)]
\item Flux density contours (isolines $S_\nu$) - among all the contours, the most important are the borders: outer one $S_{\nu,\mathrm{min}}$ and inner one $S_{\nu,\mathrm{max}}$ (in our case there are no superponed sources, so $S_{\nu,\mathrm{max}}$ would be just the maximal observed $S_\nu$),
\item Flux density profiles: we draw profile at some declination, in chosen range of right ascensions to find out what would be min and max values observed $S_\nu$, and
\item 3D profiles - with this we can also estimate the exact area.
\end{enumerate}

\begin{figure*}[ht!]
\centering
\includegraphics[width=\textwidth]{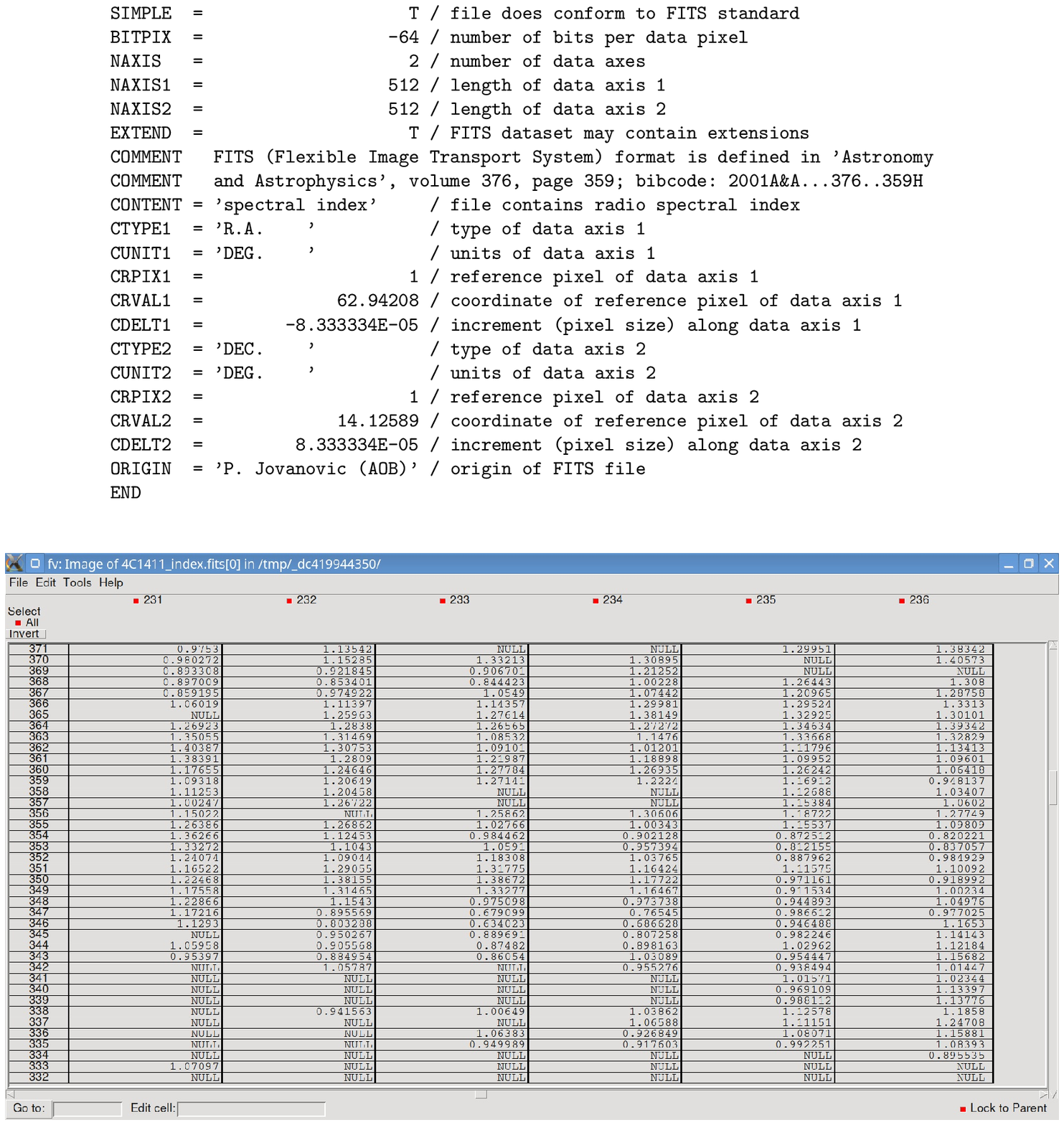}
\caption{Header unit (top) and a part of data unit (bottom) of the FITS file ''4C1411\_index.fits.gz'' containing the radio spectral index of 4C 14.11. This FITS file is available as a supplement to this paper.}
\label{fig05}
\end{figure*}

\section{Spectral index distribution}

If over some finite frequency range we can describe the amount of flux density $S_\nu$ as a function of frequency $\nu$ by the formula:

\begin{equation}
S_\nu \sim \nu^{-\alpha},
\label{equ01}
\end{equation}

\noindent where $\alpha$ is a constant, called the 'radio spectral index', we say that the flux has a 'power law dependence' on frequency (note that if flux does not follow a power law in frequency, the spectral index itself is a function of frequency). Let us emphasize here that $\alpha$ represents a quantity characterizing the spectra of continuum radio sources, defined by the formula (\ref{equ01}), while $S_\nu$ represents a measure of the strength of the radiation from a source. In radio astronomy, observed flux densities are usually extremely small and therefore are often expressed in units of the Jansky (Jy), where 1 Jy = $10^{-26}$ W m$^{-2}$ Hz$^{-1}$. By taking logarithm of (\ref{equ01}), we get:

\begin{equation}
\log S_\nu = \log K - \alpha \log \nu,
\label{equ02}
\end{equation}

\noindent where $K$ is constant. Then we determine flux density spectral index by fitting this equation to the data ($\nu$,$S_\nu$), i.e. using data for at least two frequencies, we can calculate it as negative value of slope of the line. In the brightness temperature case, we have: $T_b \sim \nu^{-\beta}$, and the brightness temperature spectral index $\beta$ can be determined in the similar way. If we express the flux density $S_\nu$ via surface brightness $B_\nu$: $S_\nu = B_\nu\ \Omega$ (where $\Omega$ is the solid angle), and for surface brightness it holds $B_\nu = (2k \nu^2 / c^2) T_b$ (where $k$ is the Boltzmann constant and $c$ the speed of light), we obtain $S_\nu = K_2 \cdot \nu^{2 - \beta}$. Then, by taking logarithm of this expression and after differentiating it over $\nu$, as well as Eq. (\ref{equ02}), we have: $\beta = \alpha + 2$.

The radio spectral index is a powerful tool for classifying cosmic radio sources and understanding the origin of the radio emission: for $\alpha < 0$ mechanism of radiation is thermal (thermal emission depends only on the temperature of the emitter), while for $\alpha > 0.1$ it is non-thermal (non-thermal emission does not depend on temperature).

The area with 4C 14.11 radio source, in equatorial coordinates, is given in Figure \ref{fig02}. We give the flux density across this radio source, at the two frequencies: 1450 MHz (left) and 8440 MHz (right), while the surrounding is denoted with white color. The corresponding flux density scales are given (in Jy). It is a FR II type of radio galaxy. It as a triple source, with a radio core, and two radio lobes located on opposite sides. Both lobes contain multiple hot spots represented by peaks in flux density distribution. Intensity of these peaks are asymmetrical because southern hot spots are more pronounced then northern in both frequencies.

With aim to check how we resolved the source from the surrounding, we also give the flux densities over its area in the form of 3D plot (see Figure \ref{fig03}). From Figs. \ref{fig02} and \ref{fig03} it is obvious that AGN at 8440 MHz has much more pronounced peak compared to 1450 MHz, but for lobs and their hot spots situation is opposite.

Figure \ref{fig04} shows spectral index map in case of 4C 14.11 radio galaxy between 1450 MHz (21 cm) and 8440 MHz (3.6 cm). The final spectral index image was obtained by combining the flux distributions of the regions detected at both frequencies (see Figure \ref{fig02}). Here, we would like to point out that the spectral index map for this radio source, over its complete area, is presented for the first time. Over the inner parts of both lobes, spectral index $\alpha$ is usually flat and significantly higher than 1.2. Spectral index over the outer parts of the lobes is relatively steep. The spectral index in the center of AGN has value $\alpha \approx$ -0.3 and denote thermal radiation. Besides AGN, over the area of the lobes spectral indices are between $\alpha \approx$ 0.7 and 2.1, which denotes non-thermal (synchrotron) radiation.

The numerical values of the spectral index presented in Fig. \ref{fig04} are also provided in the form of FITS file ''4C1411\_index.fits.gz'', which is available as a supplement to this paper. The header unit, as well as the part of data unit (table) of this FITS file are presented in Fig. \ref{fig05}.

\section{Conclusions}

There are only few papers where analyses of radio source 4C 14.11 is presented. Radio source 4C 14.11 is one of the 17 sources analyzed in paper by \citet{hard97}. Authors presented both low resolution (2$''$.40) and high resolution (0$''$.23) maps of 4C 14.11 at 8.4 GHz, where several hot spots were reported in both lobes (see Figures 5, 6 and 7 in mentioned paper). In paper by \citet{fern19} 20 radio sources at 8.4 GHz are analyzed, and one of them is 4C 14.11. In his analysis the author also used data from \citet{hard97}. In Figure 1(a), a total intensity contour map with 0$''$.83 resolution is presented. The author observed multiple features in both lobes. In paper by \citet{rudn02} a simple method is explored for examining the structures and spectral index distributions on different scales in Galactic and extragalactic radio sources, based on a multiresolution filtering technique. The data for radio source 4C 14.11 are taken from \citet{hard97}. Author briefly discusses its spectral behavior and identifies ridge seen in the southeast with a jet, based on the high-resolution map which shows that one component is a linear feature aligned with the core. Here we demonstrated that the most of these reported features are also present in our spectral index map shown in Fig. \ref{fig04}. Therefore, our results enable one to determine the dominant mechanism of radiation in each of these regions, using the corresponding local value of the spectral index from Fig. \ref{fig04}.

For DRAGN 4C 14.11 we obtained flux densities and variation of spectral indices over its area using available frequency surveys at 1450 MHz (21 cm) and 8440 MHz (3.6 cm). We wish to emphasize that we calculated the distribution of radio spectral index over the complete area of this source and presented it graphically and numerically (in FITS data format), for the first time.

At the both frequencies, the southern lobe has greater flux density than northern one, and also the both jets are weak. The hot spots of both lobes have greater flux density than the AGN at 1450 MHz, while in case of 8440 MHz the situation is opposite. Distinction between hot spots and rest of the lobes are much smaller in the spectral index than in the flux. The spectral index map (Figure \ref{fig04}) shows a clear steepening moving from the inner regions of the lobes towards the edges of the lobes. Over the inner parts of both lobes, spectral index $\alpha$ is flat in average and significantly higher than 1.2, indicating that 4C 14.11 is old AGN. Besides, steeper spectrum over the outer parts of the lobes indicates the greater age of the particles in the plasma. These results are consistent with the findings of \citet{hard97}.

Therefore, the obtained spectral index distribution can be helpful for a better understanding of the evolution of 4C 14.11 radio source. We found that the non-thermal (synchrotron) radiation dominates over the areas of the lobes of this FR II source.

\section*{Acknowledgments}
This work is supported by Ministry of Education, Science and Technological Development of the Republic of Serbia. PJ wishes to acknowledge the support by this Ministry through the project contract No. 451-03-68/2022-14/200002.

\bibliographystyle{model5-names}
\biboptions{authoryear}
\bibliography{literatura}

\end{document}